# Toward Integrated Microwave Diagnostics for Future Fusion Devices: Design of a Compact Waveguide Diplexer

Xiaoliang Li[1,a)], Valerian Hall-Chen[1], Umesh Kumar[1], Eddie Chua[1], Tengfei Tang[1], Andy Yeoh[1], Khadija Rahman[2], Xinhang Xu[2], and Yilun Zhu[2]

[1]*Future Energy Acceleration & Translation (FEAT), Agency for Science, Technology and Research (A*STAR), 2 Fusionopolis Way, Innovis #08-03, Singapore 138634, Republic of Singapore.*
[2]*University of California Davis, Davis, California, 95616, USA*
[a)]Author to whom correspondence should be addressed: Li_Xiaoliang@a-star.edu.sg

Microwave diagnostics are essential for plasma control and physics studies in magnetic confinement fusion, providing measurements of electron density, electron temperature, and energetic-particle behavior. However, diagnostic access in future burning-plasma reactors will be severely constrained by breeding blankets, neutron shielding, and compact reactor geometries, requiring reduced blanket opening areas while maintaining diagnostic capability. This work presents a compact waveguide diplexer for front-end integration of multiple microwave diagnostics. The proposed diplexer combines a waveguide T-junction with two iris-coupled sixth-order Chebyshev bandpass filters, enabling microwave reflectometry and electron cyclotron emission (ECE) diagnostics to share a common transmission line while operating in separate frequency bands. As an initial study, the diplexer is designed for the 65–70 GHz reflectometry and 85–90 GHz ECE systems on the J-TEXT tokamak. Full-wave electromagnetic simulations demonstrate insertion losses below 1 dB and channel isolation exceeding 40 dB across both operating bands. The proposed methodology provides a compact and scalable approach to microwave diagnostic integration and establishes a practical basis for multi-diagnostic front-end systems in future fusion devices.

## I. Introduction

Microwave diagnostics[1][2] have become indispensable tools for magnetic confinement fusion research by providing non-perturbative measurements of a wide range of plasma parameters, including electron density[3][4][5], electron temperature[6][7][8], magnetic fluctuations[9], plasma turbulence[10][11][12], and energetic particle distributions[13]. Diagnostic techniques such as microwave reflectometry, electron cyclotron emission (ECE)[14][15], Doppler backscattering (DBS)[16][17], interferometry, and collective Thomson scattering (CTS)[18][19] have been extensively implemented on present-day tokamaks and stellarators to investigate plasma transport, magnetohydrodynamic instabilities, edge physics, and plasma heating. Continuous developments in millimeter-wave instrumentation, integrated receiver technologies[20], synthetic diagnostics[21][22][23], and real-time data analysis[24] have further enhanced the capability of microwave diagnostics, making them essential components of modern fusion experiments.

While remarkable progress has been achieved on existing experimental devices, future burning plasma reactors will impose substantially more stringent engineering constraints on diagnostic systems. Compared with present-day tokamaks, reactor-scale devices are expected to provide substantially fewer diagnostic access ports because of the installation of breeding blankets, neutron shielding structures, and increasingly compact reactor geometries[25]. At the same time, future plasma operation will require simultaneous measurements of multiple plasma parameters with higher spatial and temporal resolution to support advanced plasma control and machine protection. Consequently, the conventional approach of deploying independent transmission lines and dedicated diagnostic windows for individual microwave diagnostics will become increasingly impractical. Developing compact front-end integration technologies capable of combining multiple microwave diagnostics into a shared transmission system therefore represents an important technological challenge for future fusion reactors.

Several approaches have been explored to improve the utilization of diagnostic access ports as shown in Table I. Quasi-optical beam splitters have been widely employed to combine multiple microwave diagnostics because of their relatively simple implementation[26]. However, beam splitters inevitably divide the incident power between transmission and reflection channels, resulting in significant power loss that is particularly undesirable for weak signal detection diagnostics such as ECE and CTS. More recently, frequency-selective surfaces (FSSs) have been introduced to improve transmission efficiency by providing frequency-dependent signal separation[27]. Although these approaches have been successfully applied to several existing fusion devices, they generally rely on large quasi-optical assemblies consisting of mirrors, lenses, beam splitters, and free-space propagation paths. Such systems are mechanically complex, sensitive to alignment errors, occupy considerable installation space, and remain vulnerable to electromagnetic interference, making them less attractive for future reactor environments where compactness, robustness, and long-term operational reliability are essential.

*Table I. Comparison of conventional approaches for integrated microwave diagnostics.*

| Item | Quasi-optical beam splitters [26] | Frequency-selective surfaces (FSSs) [27] |
|---|---|---|
| Operating principle | Divide the incident microwave power into transmitted and reflected components to combine multiple diagnostic channels | Provide frequency-selective transmission and reflection to separate microwave signals operating at different frequencies |
| Key advantages | Simple implementation and compatibility with established quasi-optical diagnostic systems | Frequency-selective operation and improved separation of multiple microwave channels |
| Power efficiency | Intrinsically limited by power division between transmission and reflection paths; reduced available signal power can be undesirable for sensitive passive diagnostics such as ECE | Can provide frequency-selective signal routing without direct power splitting; however, performance depends on the FSS design and associated quasi-optical components |
| System complexity | Requires free-space propagation and multiple optical components, such as mirrors and lenses | Typically requires free-space propagation together with supporting quasi-optical components and alignment structures |
| Integration limitations | Relatively large installation footprint; sensitive to alignment and mechanical displacement | Relatively large and mechanically complex quasi-optical assemblies; alignment and packaging can become challenging |
| Reactor-scale integration | Limited by power efficiency, system volume, and alignment sensitivity | Limited by system volume, mechanical complexity, and alignment requirements |

Waveguide diplexer approach provides an attractive alternative for microwave diagnostic integration. Compared with quasi-optical systems, waveguide components offer lower transmission loss, superior electromagnetic shielding, higher mechanical stability, and significantly reduced installation volume. More importantly, frequency-selective waveguide components enable signal routing directly within enclosed transmission structures without introducing additional quasi-optical elements. Waveguide diplexers have been extensively developed for communication and radar applications. In fusion applications, however, microwave diagnostic integration has mainly relied on quasi-optical beam splitters and frequency-selective surfaces. The application of a compact waveguide diplexer for sharing a common transmission path between multiple fusion microwave diagnostics remains comparatively unexplored. The novelty of this work lies in the system-level application and optimization of a waveguide diplexer for integrated fusion microwave diagnostics, with simultaneous consideration of channel isolation, transmission loss, compactness, and diagnostic compatibility.

In this work, a compact waveguide diplexer is proposed as the enabling component for integrated microwave diagnostics on fusion devices. The proposed architecture combines two iris-coupled sixth-order Chebyshev waveguide bandpass filters through a waveguide T-junction, allowing microwave reflectometry and electron cyclotron emission diagnostics to share a common front-end transmission line while operating over independent frequency bands. As an initial proof-of-concept, the diplexer is designed for the existing 65–70 GHz reflectometry and 85–90 GHz ECE systems on the J-TEXT tokamak. These frequency bands are particularly relevant to the pedestal region, enabling simultaneous, co-located measurements of electron density and electron temperature, which are valuable for studying edge plasma physics. A complete design methodology, including system specification, filter synthesis, electromagnetic optimization, and diplexer integration, is developed and evaluated using full-wave electromagnetic simulations. Although the present study focuses on the J-TEXT operating frequencies, the proposed methodology is readily scalable to other frequency bands and provides a general framework for future multi-diagnostic microwave integration on reactor-scale fusion devices.

The remainder of this paper is organized as follows. Section II presents the overall architecture of the proposed integrated microwave diagnostic system together with the selection of operating frequency bands. Section III describes the design methodology of the iris-coupled waveguide filters and the diplexer configuration. Section IV presents the optimization procedure and simulated performance of the individual filters and the complete diplexer. Finally, Section V summarizes the present work and discusses future experimental implementation on the J-TEXT tokamak as well as extensions toward integrated microwave diagnostic systems for future burning plasma reactors.

## II. Integrated Microwave Diagnostic Architecture

The proposed solution is based on front-end integration of multiple microwave diagnostics through a frequency-selective waveguide architecture. Figure 1 illustrates the proposed integrated microwave diagnostic architecture. The system consists of a waveguide diplexer connected to a shared front-end transmission line, enabling multiple microwave diagnostics to utilize a common antenna and diagnostic window. In the present study, the diplexer integrates the reflectometry and ECE diagnostics, although the same architecture can be readily extended to additional microwave diagnostics, such as DBS, interferometry, and CTS.

The proposed diplexer adopts a passive frequency-selective routing scheme. Microwave signals entering the common port are routed to the corresponding diagnostic channels according to their frequency bands. As a result, both diagnostics can operate simultaneously while sharing the same front-end transmission system. Compared with conventional standalone microwave diagnostic systems, the proposed architecture significantly reduces the required front-end hardware and diagnostic access while maintaining independent signal paths for each diagnostic. The compact waveguide implementation also provides excellent mechanical stability and electromagnetic shielding, making it well suited for future integrated microwave diagnostic systems.

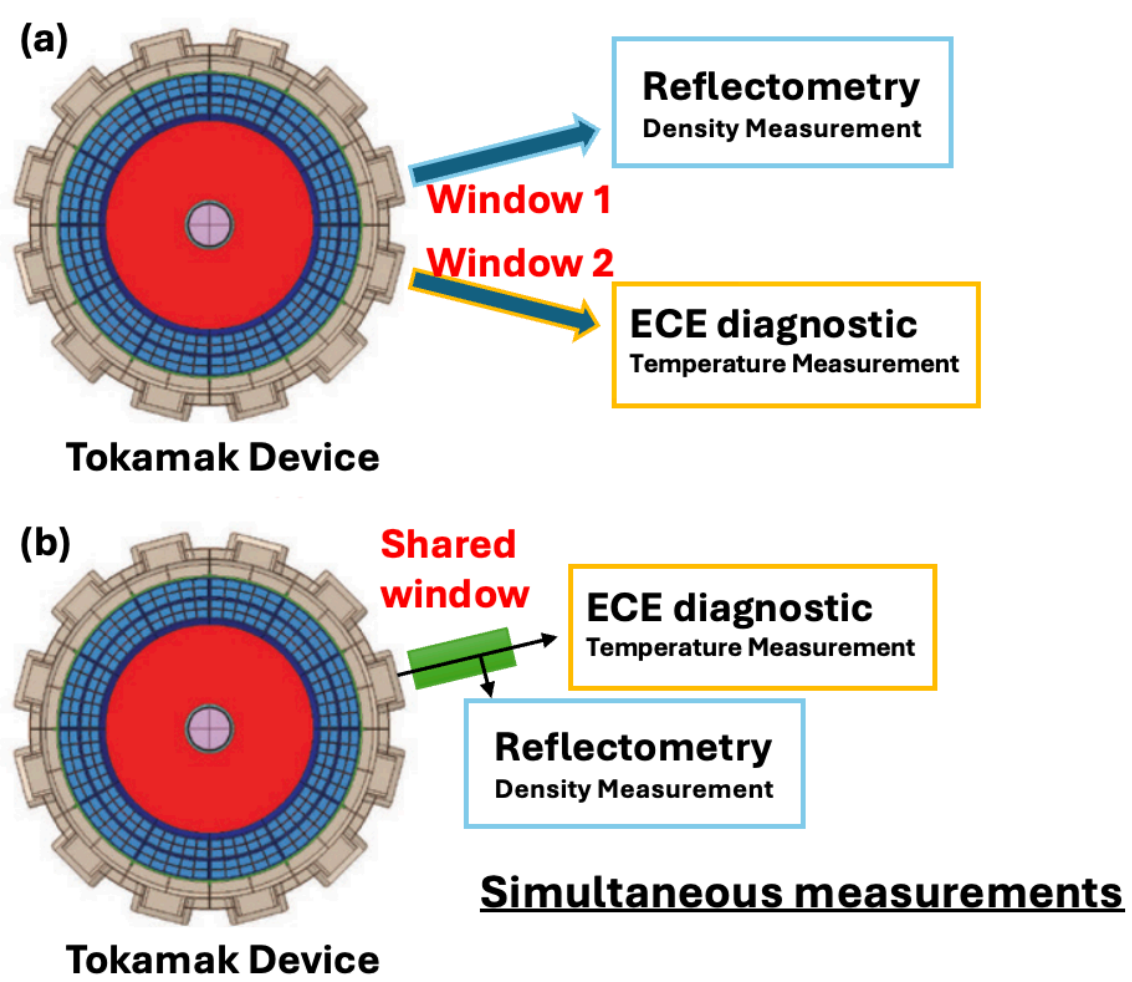

*Figure 1. Concept of an integrated microwave diagnostic architecture for fusion devices. (a) Conventional configuration using two separate windows for reflectometry and ECE diagnostics. (b) Proposed configuration using a shared diagnostic window and an integrated waveguide front-end for simultaneous density and temperature measurements.*

To experimentally demonstrate the proposed integration concept, the diplexer is designed for the existing microwave diagnostic systems on the J-TEXT tokamak [28]. The lower-frequency channel covers 65–70 GHz for microwave reflectometry [29], while the higher-frequency channel covers 85–90 GHz for electron cyclotron emission (ECE) [30]. These frequency bands are selected to ensure direct compatibility with the existing J-TEXT transmission lines and diagnostic hardware, enabling experimental validation without modification of the launchers, receivers, or plasma-facing components. The 15 GHz separation between the two bands also provides sufficient frequency spacing for high channel isolation while enabling practical iris-coupled Chebyshev filter designs. Both bands fall within the WR-12 waveguide range (60–90 GHz), allowing the complete diplexer to be implemented using a common waveguide size and compact housing without additional waveguide transitions. Moreover, the relatively narrow fractional bandwidths of the two channels, approximately 7.4% and 5.7%, respectively, are well suited to high-order Chebyshev filter synthesis, facilitating low insertion loss and sharp roll-off. These characteristics make the selected frequency bands a suitable test case for demonstrating the proposed integrated microwave diagnostic architecture and future design will expand the bandwidth.

Importantly, these operating frequencies are determined by the existing J-TEXT diagnostic systems rather than by the proposed diplexer architecture. The proposed design methodology is frequency independent and can be readily scaled to other millimeter-wave and sub-terahertz frequency bands through appropriate scaling of the waveguide dimensions and resonator parameters. Consequently, the present design represents a practical proof-of-concept for future integrated microwave diagnostics rather than a frequency-specific solution.

Figure 2 shows the proposed waveguide diplexer configuration. The device adopts a three-port waveguide T-junction topology consisting of one common port and two branch ports. The common port is connected to the shared transmission line and antenna system, while each branch port is connected to an iris-coupled waveguide bandpass filter corresponding to one diagnostic channel. Each channel filter is designed as a sixth-order Chebyshev waveguide bandpass filter using inductive irises and half-wavelength resonant cavities[31][32][33]. The lower-frequency filter transmits the reflectometry signals while rejecting the ECE frequency band. Conversely, the higher-frequency filter passes the ECE signals and suppresses the reflectometry frequencies.

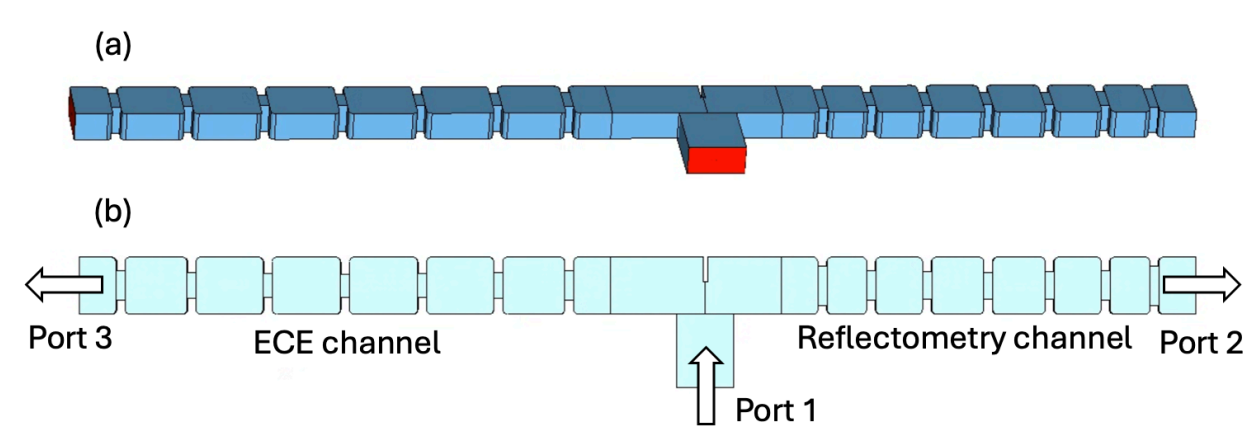

*Figure 2. Configuration of the proposed waveguide diplexer: (a) three-dimensional view and (b) two-dimensional schematic.*

When microwave signals enter the common port, the frequency-selective characteristics of the two filters automatically direct the signals toward their corresponding diagnostic channels. Signals within the reflectometry band propagate through the lower-frequency branch with minimal attenuation, while signals in the ECE band are transmitted through the higher-frequency branch. Owing to the high out-of-band rejection of the individual filters, high isolation between the two channels is achieved without employing active microwave switches or additional frequency-selective components.

To ensure compatibility with the existing J-TEXT microwave diagnostic systems, several performance requirements are specified for the proposed diplexer. The insertion loss of each channel is required to be less than 1 dB within the operating band to minimize signal attenuation. The channel isolation is specified to be greater than 30 dB to suppress mutual interference between the reflectometry and ECE channels. These requirements serve as the design objectives for the subsequent filter synthesis and electromagnetic optimization.

## III. Design Methodology

The overall design workflow adopted in this work is summarized in Figure 3. The proposed diplexer consists of two iris-coupled waveguide bandpass filters connected through a waveguide T-junction. Rather than directly optimizing the complete diplexer, a sequential design methodology is adopted to improve optimization efficiency while maintaining desired transmission performance.

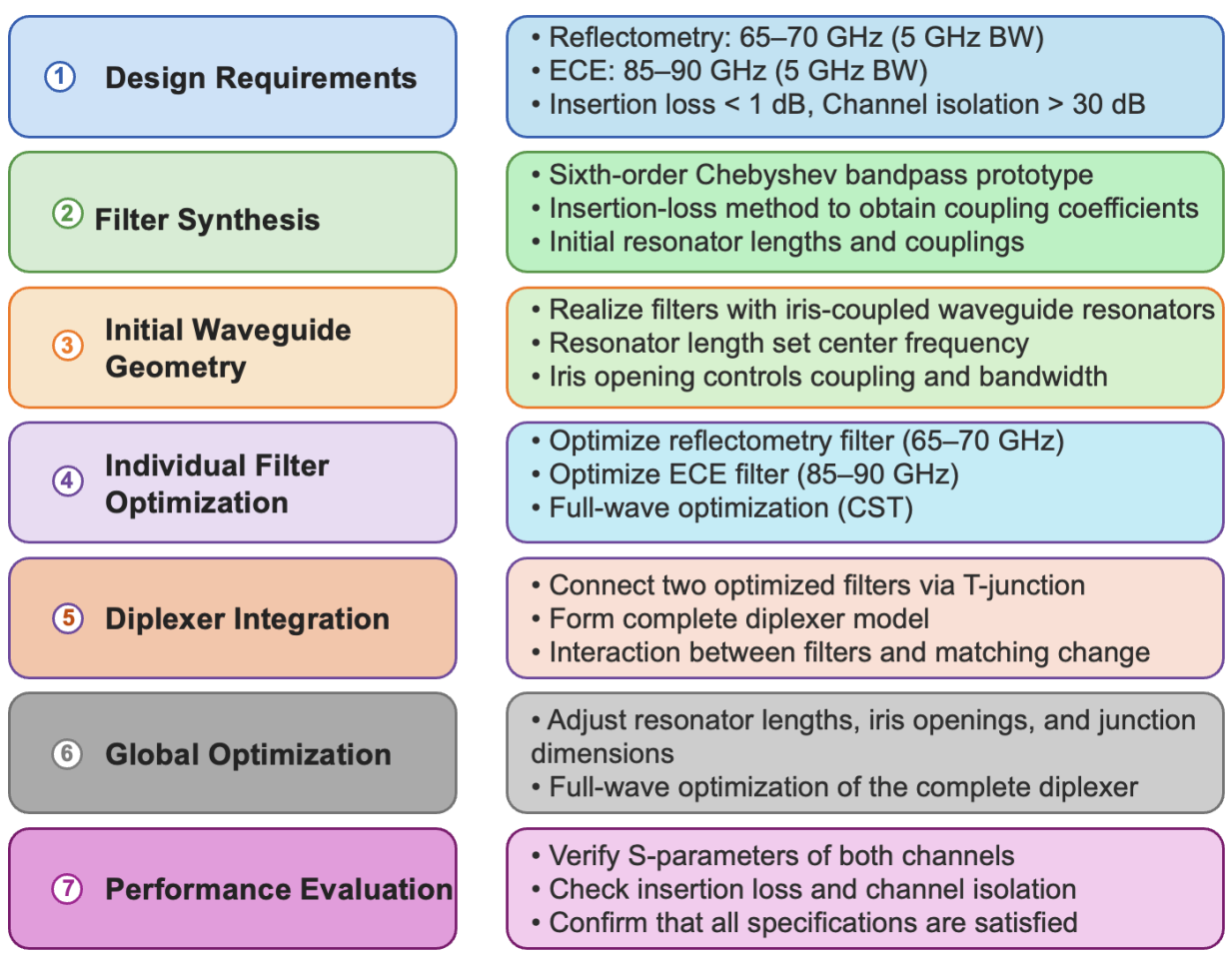

*Figure 3. Design workflow of the proposed waveguide diplexer.*

The design begins with the synthesis of two waveguide bandpass filters according to the operating frequencies and bandwidths of the reflectometry and ECE systems. Based on the theoretical synthesis described in [34], a sixth-order Chebyshev bandpass filter is sufficient to meet the above performance requirements, providing a good trade-off between passband flatness, out-of-band rejection, and fabrication complexity[35]. Compared with Butterworth filters, Chebyshev filters exhibit a much steeper roll-off between the passband and stopband while maintaining an acceptable level of passband ripple. This characteristic enables high channel isolation with a relatively low filter order, making the Chebyshev response particularly suitable for waveguide diplexer applications. Classical insertion-loss synthesis is then employed to determine the filter order, coupling coefficients, and resonator electrical lengths[34]. The normalized low-pass prototype is subsequently transformed into the corresponding bandpass filter, from which the initial physical dimensions of the resonators and coupling irises are obtained.

The synthesized filters are implemented using iris-coupled rectangular waveguide resonators, as illustrated in Figure 4. In this configuration, each resonator operates near its fundamental resonance mode, while adjacent resonators are electromagnetically coupled through inductive irises. The resonator lengths primarily determine the center frequencies of the filters, whereas the iris openings control the coupling strength between neighboring cavities and therefore the filter bandwidth. To reduce the number of optimization variables and simplify fabrication, both channel filters are designed using symmetric six-resonator configurations. The analytical dimensions obtained from the filter synthesis provide the initial geometry for the subsequent full-wave electromagnetic simulations.

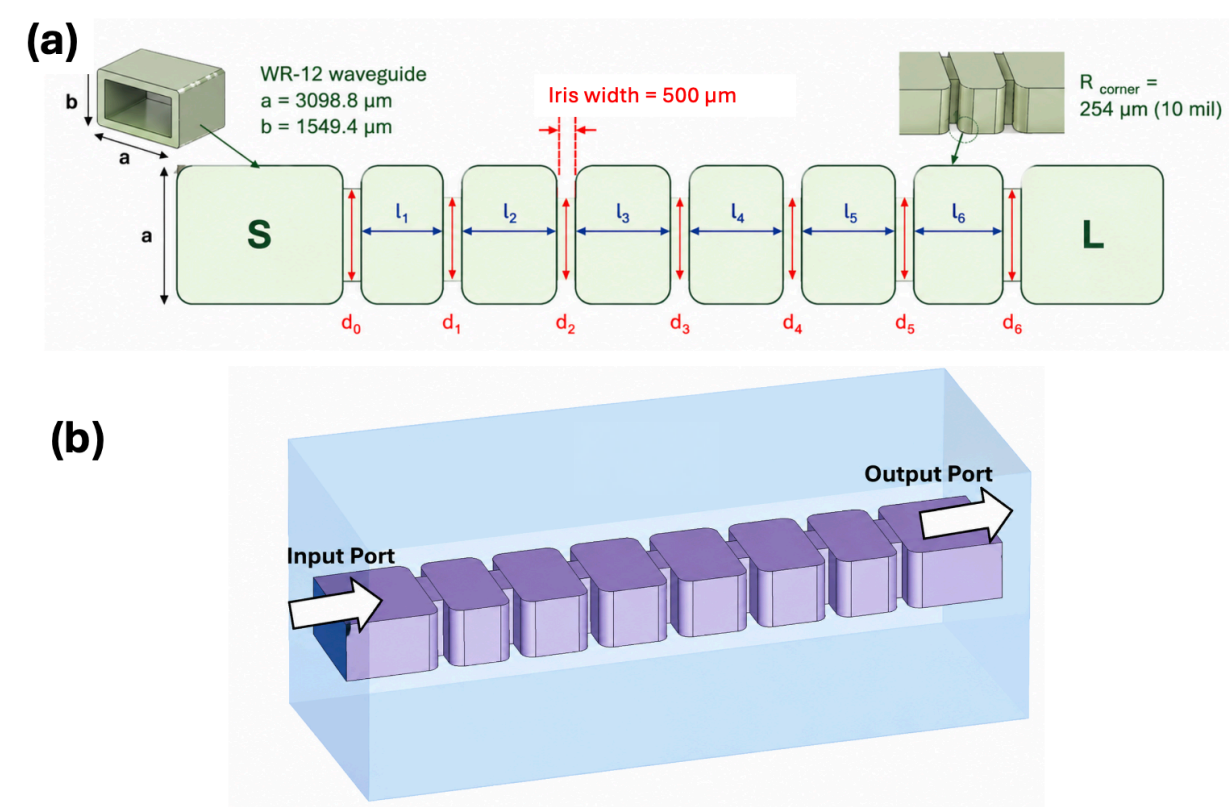

*Figure 4. Geometry of the iris-coupled sixth-order Chebyshev bandpass filter. (a) Two-dimensional schematic showing the filter geometry and design parameters. (b) Three-dimensional view of the filter structure.*

Since the equivalent circuit model cannot accurately account for higher-order mode coupling, conductor losses, waveguide discontinuities, and the electromagnetic interaction introduced by the waveguide T-junction, full-wave electromagnetic optimization is required to obtain the final diplexer geometry. In the full-wave simulations, the waveguide walls were modeled as perfect electric conductors, and adaptive mesh refinement was employed in CST Microwave Studio to ensure sufficient numerical accuracy. The optimization is performed in two stages. First, the key geometric parameters are manually adjusted to bring the filter response close to the target specifications. The resulting geometry is then used as the initial solution for the built-in CST Microwave Studio[36] optimization algorithm, which automatically refines the design parameters to achieve the desired passband insertion loss and return loss while maintaining the target operating frequencies. The optimization is carried out sequentially. First, the two channel filters are optimized independently and then integrated through the waveguide T-junction to establish the complete diplexer model.

Direct integration of the two optimized filters, however, does not produce the desired diplexer response. The introduction of the T-junction modifies the electromagnetic boundary conditions of both filters, resulting in noticeable frequency shifts, impedance mismatches, and degradation of channel isolation. Therefore, a global optimization of the complete diplexer is subsequently performed to compensate for these interactions. The resonator lengths, iris openings, and junction dimensions are iteratively adjusted until the insertion loss, return loss, and channel isolation simultaneously satisfy the design requirements.

## IV. Design optimization and iteration improvements

To evaluate the proposed waveguide diplexer, full-wave electromagnetic simulations were performed using CST Microwave Studio[36] following the design methodology described in Section III. The simulated performances of the individual channel filters and the complete diplexer are presented in this section.

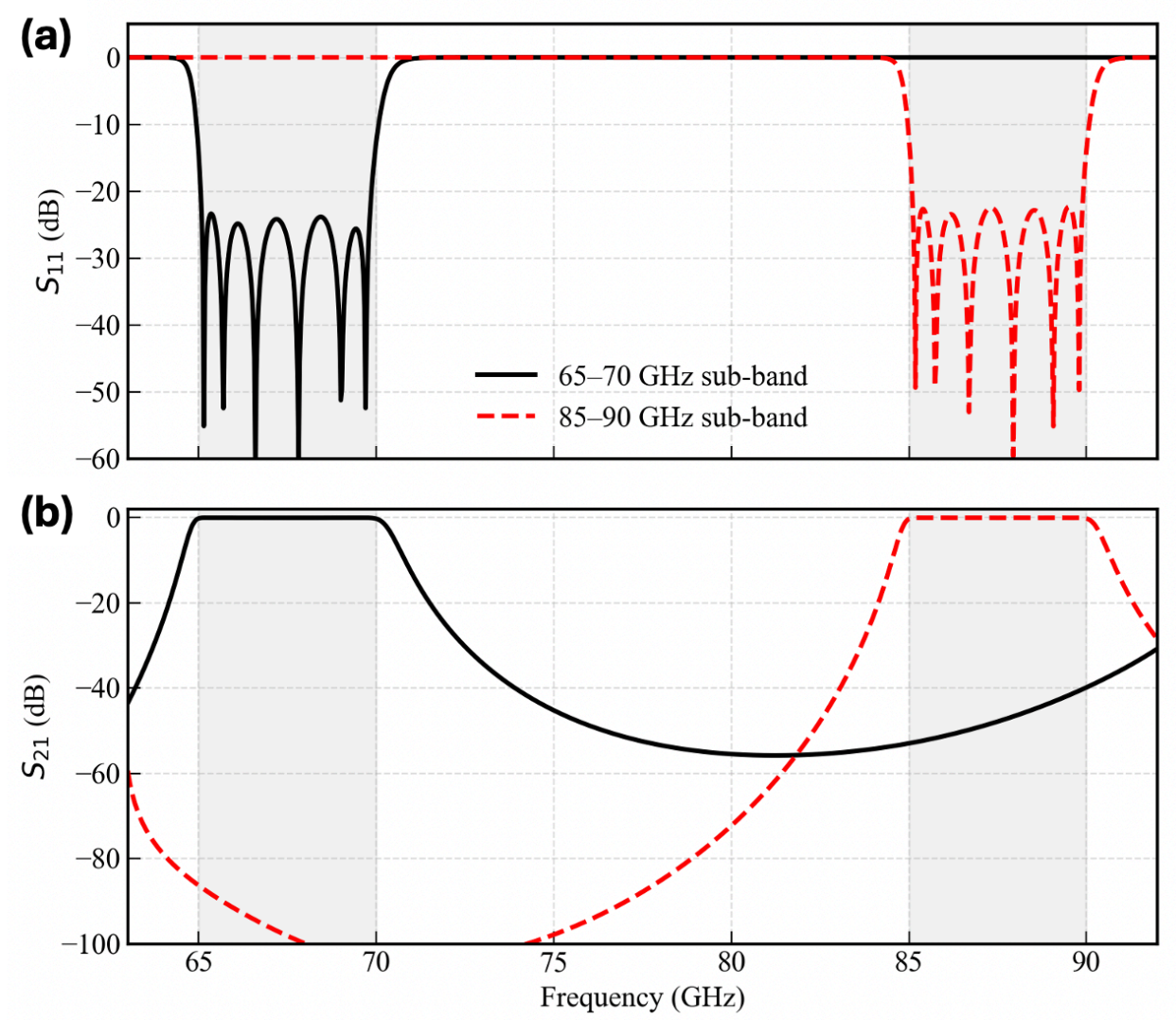

*Figure 5. Simulated S-parameters of the two optimized channel filters. (a) S11. (b) S21. The solid black and dashed red curves represent the 65-70 GHz reflectometry channel filter and the 85-90 GHz ECE channel filter, respectively.*

Figure 5 shows the simulated responses of the two independently optimized channel filters. The 65–70 GHz filter achieves an insertion loss below 1 dB across the reflectometry band, while the 85–90 GHz filter exhibits similarly low insertion loss over the ECE band. Both filters also show steep out-of-band rejection, providing the frequency selectivity required for effective channel separation in the diplexer. The corresponding optimized dimensional parameters are summarized in Table II.

*Table II. Dimensional parameters of the independently optimized channel filters.*

| Parameter (μm) | Reflectometry Channel (65-70 GHz) | ECE Channel (85-90 GHz) |
|---|---|---|
| $l_1 = l_6$ | 2149 | 3335 |
| $l_2 = l_5$ | 2537 | 3606 |
| $l_3 = l_4$ | 2626 | 3671 |
| $d_0 = d_6$ | 2076 | 1798 |
| $d_1 = d_5$ | 1606 | 1389 |
| $d_2 = d_4$ | 1476 | 1277 |
| $d_3$ | 1453 | 1258 |
| S/L | 2000 | 2000 |

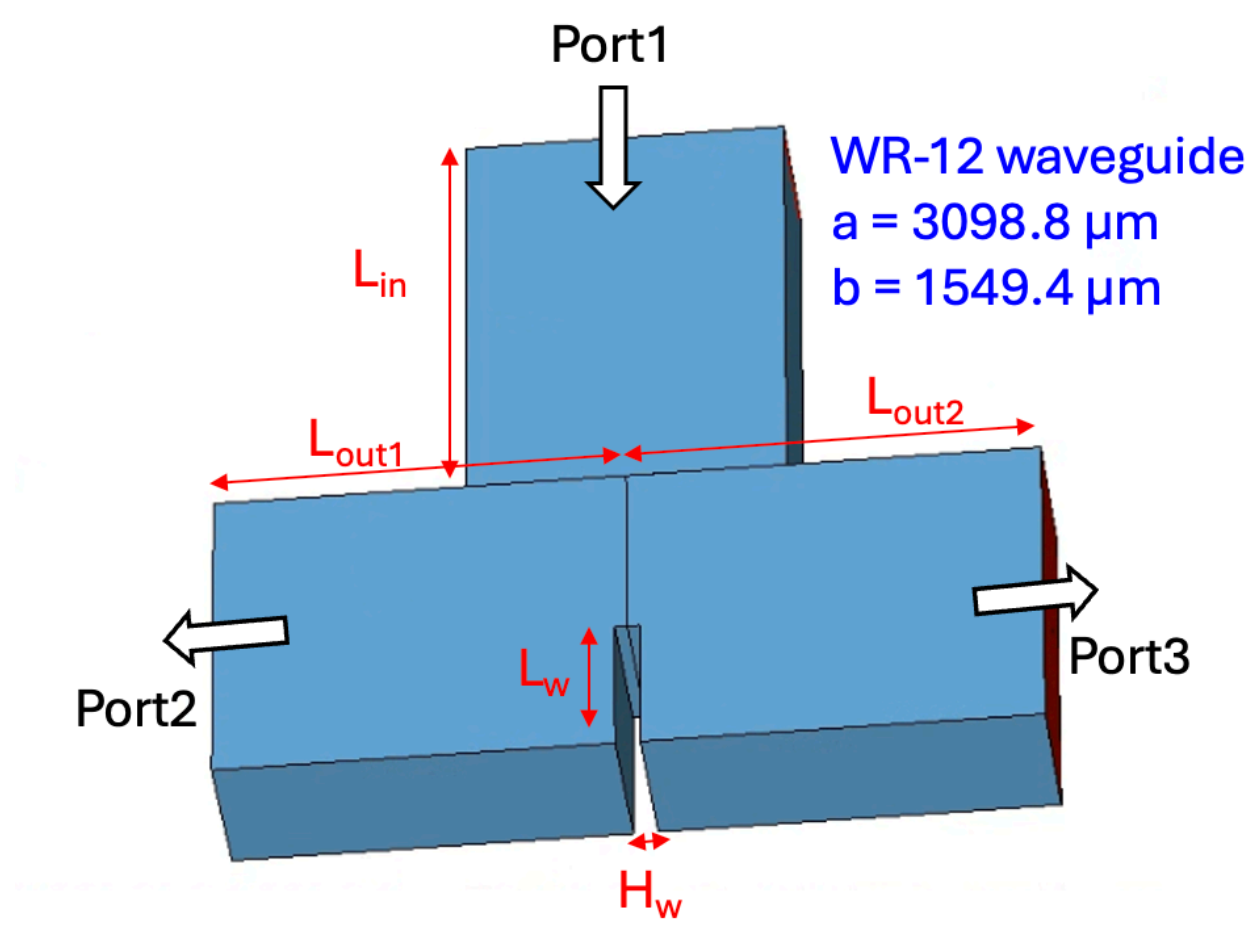

*Figure 6. 3D structure of the T-junction.*

*Table III. Dimensional parameters of the independently optimized waveguide T-junction.*

| | $L_w$ | $H_w$ | $L_{out1}$ | $L_{out2}$ | $L_{in}$ |
|---|---|---|---|---|---|
| Value (μm) | 1346 | 240 | 4000 | 4000 | 4000 |

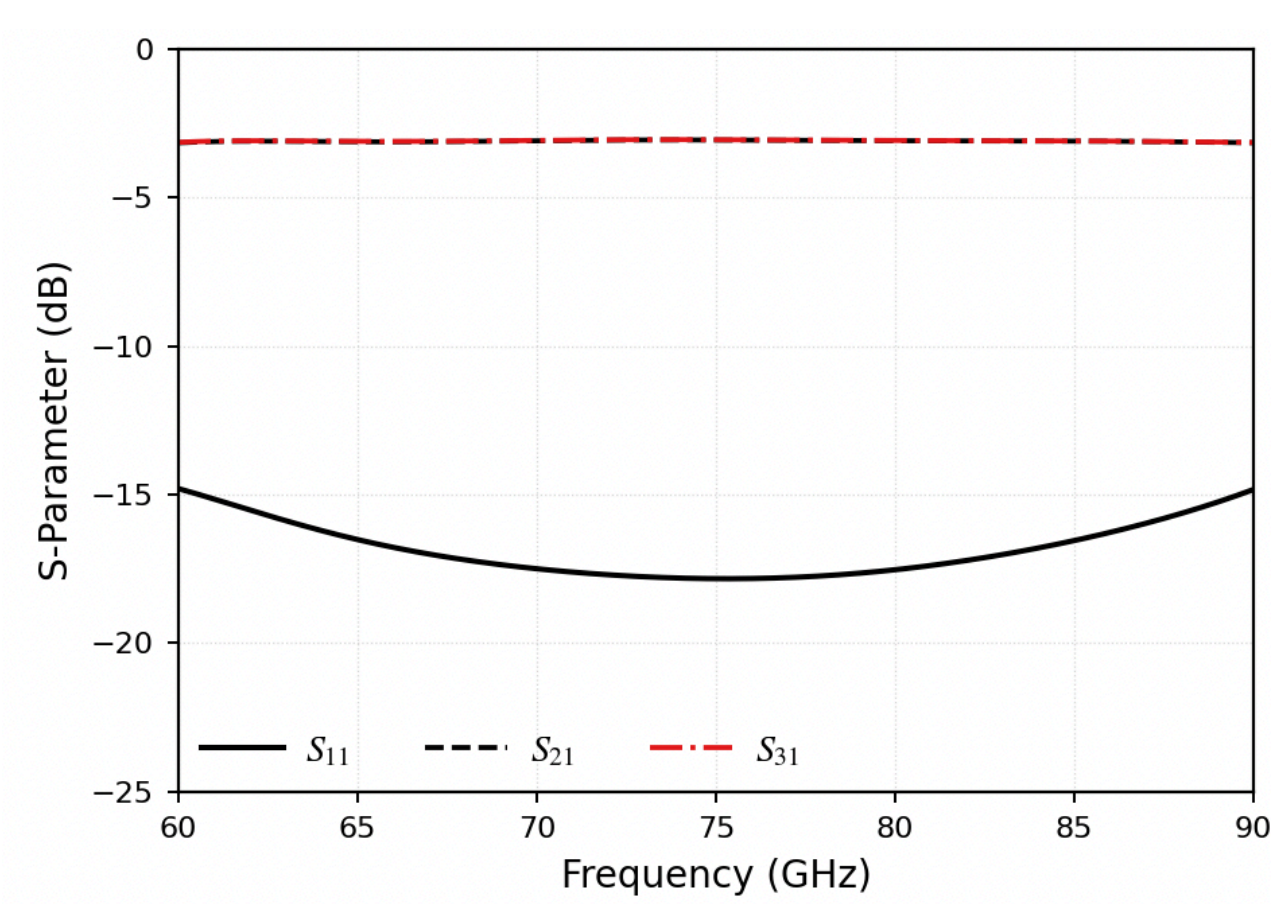

*Figure 7. Simulated S-parameters of the optimized T-junction.*

The optimized channel filters were subsequently integrated through a waveguide T-junction to form the complete diplexer. As shown in Figure 6, a wedge structure was introduced at the junction to improve the power-division performance. Initial simulations showed that directly connecting two WR-12 waveguides to form a conventional T-junction resulted in significant impedance mismatch and nonideal power division, resulting in significant impedance mismatch and an unbalanced power division between the two output ports. To address this issue, a wedge structure was introduced at the junction, and its length ($L_w$) and height ($H_w$) were systematically optimized through full-wave electromagnetic simulations. The optimized wedge provides improved impedance matching and enables a more balanced power distribution between the two output ports. The optimized dimensional parameters of the T-junction are summarized in Table III. The corresponding simulated S-parameters are presented in Figure 7, demonstrating the improved power-division performance of the optimized T-junction.

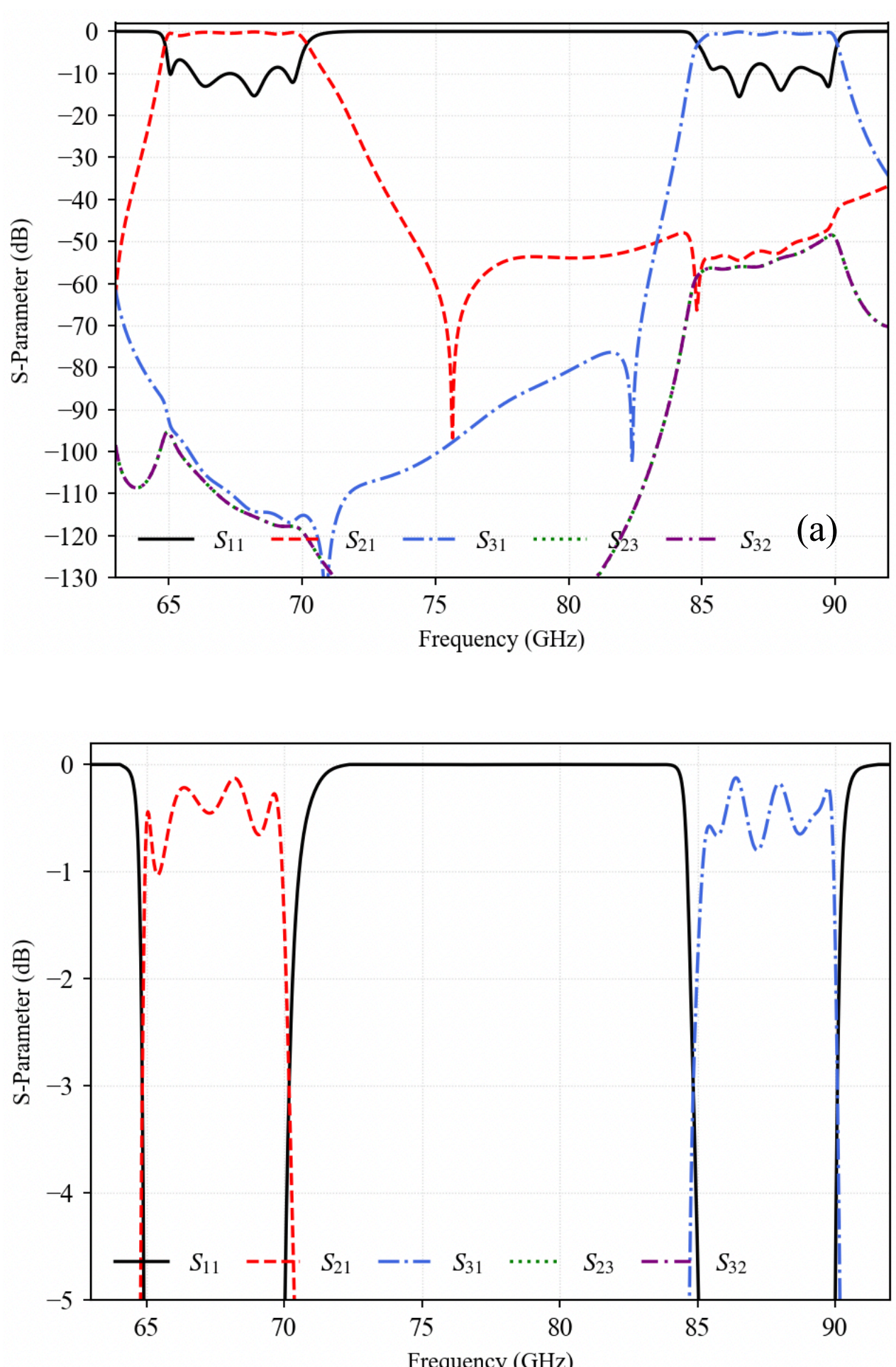

(b)

*Figure 8. Simulated S-parameters of the optimized diplexer: (a) overall response showing the channel isolation and return characteristics; (b) in-band transmission responses for the 65-70 GHz and 85-90 GHz channels, both exhibiting insertion losses below 1 dB.*

After integrating the two optimized filters with the T-junction, the overall S-parameters deviated significantly from the target specifications. This degradation indicates that the electromagnetic interaction among the individual components must be considered at the system level. Therefore, the lengths of the two T-junction output ports and selected filter parameters were jointly optimized to compensate for these interactions and improve the overall diplexer performance. The simulated S-parameters after optimization are presented in Figure 8. Desired transmission characteristics are achieved for both diagnostic channels. The insertion loss ($S_{21}$ and $S_{31}$) remains below 1 dB over both operating bands. As shown in Fig. 8, $S_{21}$ and $S_{31}$ remain above −1 dB throughout the 65–70 GHz and 85–90 GHz bands, respectively, corresponding to insertion losses below 1 dB. More importantly, the channel isolation ($S_{23}$ and $S_{32}$) is better than 40 dB, demonstrating effective suppression of mutual interference between the reflectometry and ECE channels. These results satisfy the specified insertion-loss and channel-isolation requirements and demonstrate the effectiveness of the proposed system-level optimization approach. The optimized geometric dimensions of the proposed diplexer are summarized in Table IV. The largest dimensional changes occur at the two T-junction branch lengths ($L_{out1}$ and $L_{out2}$), which are adjusted to optimize the overall diplexer response.

*Table IV. Final geometric dimensions of the jointly optimized waveguide diplexer.*

| Parameter (µm) | Reflectometry Channel (65-70 GHz) | ECE Channel (85-90 GHz) | T-junction |
|---|---|---|---|
| $l_1 = l_6$ | 2149 | 3360 | NA |
| $l_2 = l_5$ | 2537 | 3631 | NA |
| $l_3 = l_4$ | 2626 | 3696 | NA |
| $d_0 = d_6$ | 2126 | 1748 | NA |
| $d_1 = d_5$ | 1606 | 1389 | NA |
| $d_2 = d_4$ | 1476 | 1277 | NA |
| $d_3$ | 1453 | 1258 | NA |
| S/L | 2000 | 2000 | NA |
| $L_w$ | NA | NA | 1346 |
| $H_w$ | NA | NA | 240 |
| $L_{out1}$ | NA | NA | 5150 |
| $L_{out2}$ | NA | NA | 4100 |
| $L_{in}$ | NA | NA | 4000 |

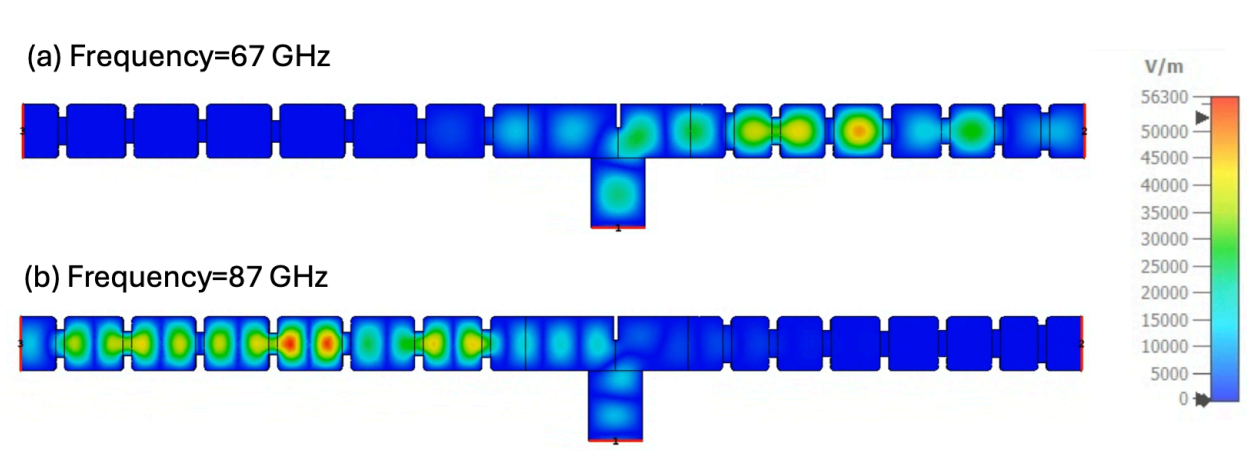

*Figure 9. Simulated electric-field distributions at representative frequencies, at (a) 67 GHz and (b) 87 GHz.*

To further illustrate the operating mechanism of the proposed diplexer, the simulated electric-field distributions at representative frequencies are presented in Figure 9. At 67 GHz, the electromagnetic field propagates almost entirely through the reflectometry branch, whereas negligible field energy is observed in the ECE channel. In contrast, at 87 GHz, the electromagnetic energy is efficiently guided into the ECE branch while the reflectometry branch is effectively isolated. The field distributions are consistent with the corresponding S-parameter responses and provide a direct visualization of the frequency-selective routing behavior.

To evaluate the robustness of the filter performance against fabrication imperfections, a tolerance analysis was performed by varying the waveguide height b, which has been identified and experimentally validated as a critical dimension in the fabrication of our high-frequency waveguide filters[32]. Based on our previous fabrication experience with high-frequency waveguide filters manufactured using high-precision computer numerical control (CNC) machining [32], the dimensional deviation of b was found to be approximately ±3 µm. To conservatively assess the fabrication tolerance, b was therefore varied from 1549.4 ± 10 µm (1539.4–1559.4 µm). As shown in Figure 10, even this relatively large dimensional variation causes no significant degradation in the filter response, with a maximum fluctuation of approximately 0.2 dB in the 85-90 GHz passband. During fabrication, the critical dimensions will be measured after each machining step, allowing the machining process to be adjusted iteratively to minimize dimensional deviations from the optimized design.

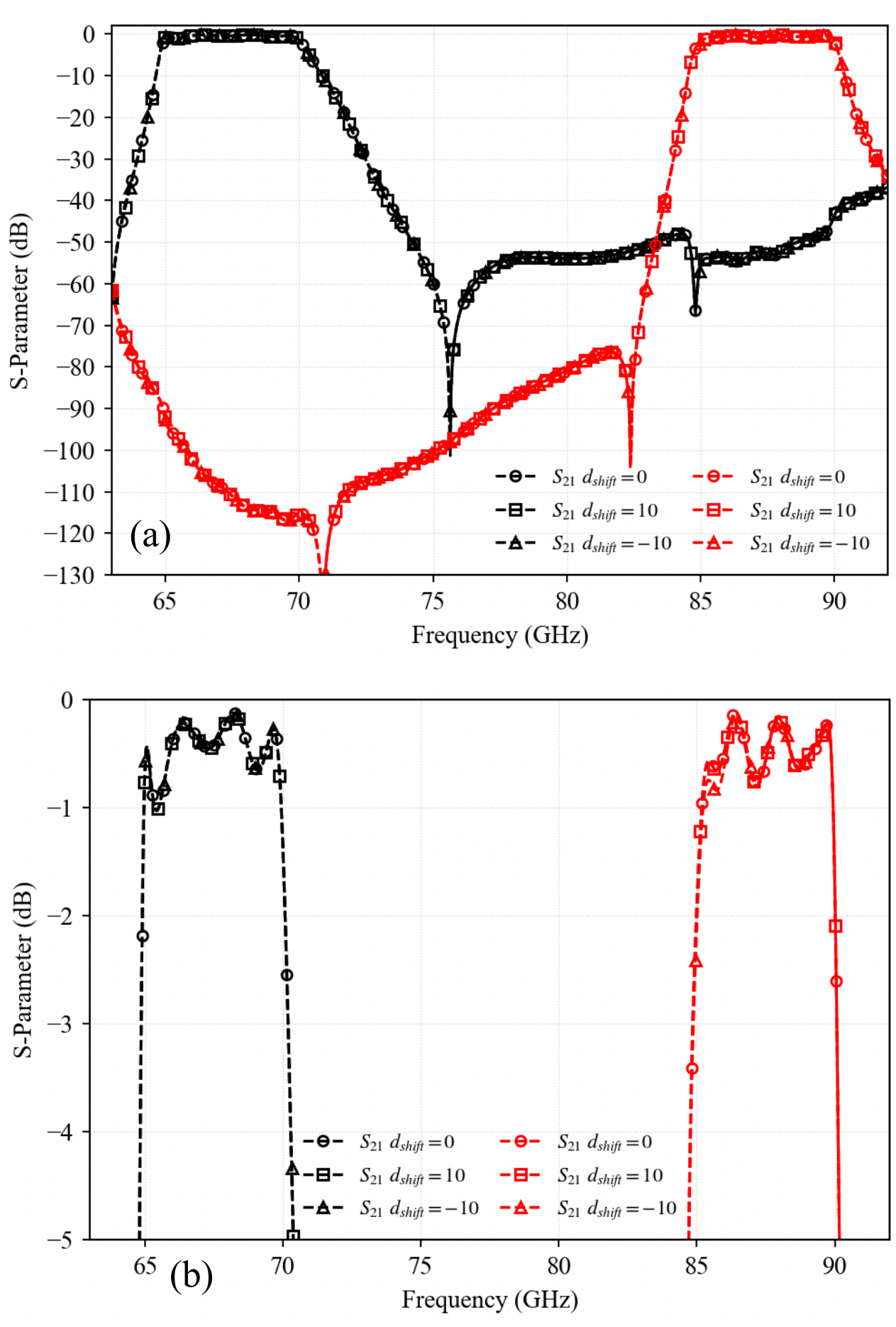

*Figure 10. Simulated tolerance analysis of the proposed diplexer for variations in the waveguide height b: (a) insertion loss of the two operating channels for b=1549.4±* 10 µm*, and (b) detailed view of the response from 0 to -5 dB.*

## V. DISCUSSION AND FUTURE WORK

The preceding analysis has primarily focused on the amplitude response of the proposed diplexer. For microwave reflectometry, however, the phase response is also important because the plasma density profile is inferred from the phase of the reflected signal. Therefore, the group delay of the 65–70 GHz reflectometry channel was evaluated from the simulated $S_{21}$ response. The group delay varies from approximately 0.25 to 0.60 ns across the operating band. This frequency-dependent delay introduces an additional phase contribution that may affect the radial localization and reconstructed density profile. During calibration, the complex transmission response of the diplexer can be characterized over the operating band using a calibrated measurement, and the corresponding frequency-dependent phase response can be incorporated into the reflectometer system transfer function for phase correction. Although the effect is expected to be small, the diplexer response should be characterized and compensated for during system calibration.

For the ECE channel, the diplexer insertion loss and its frequency dependence may affect the absolute calibration and effective receiver noise performance through signal attenuation. These effects will be quantitatively evaluated after fabrication and system integration. The diplexer will be fabricated using Aluminum Alloy 2024-T4/T6, with a conductivity of $1.734 \times 10^7$ S/m, which has been previously used and experimentally validated for our high-frequency waveguide filters. The fabricated diplexer will be experimentally characterized and integrated with the existing reflectometry and ECE systems for laboratory and plasma experiments. The proposed design methodology can be further extended to additional microwave diagnostics, such as DBS, interferometry, and CTS, providing a scalable approach for shared front-end integration in future fusion devices.

## VI. CONCLUSION

A compact waveguide diplexer has been proposed for integrated microwave diagnostics on fusion devices. The diplexer employs a waveguide T-junction combined with two iris-coupled sixth-order Chebyshev bandpass filters to integrate reflectometry and electron cyclotron emission (ECE) diagnostics through a shared front-end transmission line. A sequential design methodology, including analytical filter synthesis, electromagnetic optimization, and system-level integration, has been developed to realize the proposed architecture. Compared with conventional standalone microwave diagnostic systems, the proposed waveguide-based solution provides a compact and mechanically robust approach for front-end integration.

Full-wave electromagnetic simulations demonstrate insertion losses below 1 dB and channel isolation exceeding 40 dB over the two operating bands. The simulated electric-field distributions further verify the frequency-selective routing behavior of the diplexer. The proposed work establishes a practical basis for waveguide-based integration of multiple microwave diagnostics and provides a scalable approach toward shared front-end transmission systems for future fusion devices.

## VI. ACKNOWLEDGMENTS

This work is supported by A*STAR under Strategic Tokamak Research for Industrial Deployment, and Energy (STRIDE), Grant No: H26-MSE1527.

## VII. DATA AVAILABILITY STATEMENT

The data that support the findings of this study are available from the corresponding author upon reasonable request.

AI Declaration: ChatGPT 4.0 was used to improve the language, clarity, and readability of this manuscript. All scientific content and conclusions were developed and verified by the authors.

## VIII. REFERENCES

[1] A. Ovsyannikov, et al. Plasma diagnostics. Cambridge Int Science Publishing, 2000.

[2] N. C. Luhmann Jr, et al. "Chapter 3: microwave diagnostics." Fusion Science and Technology 53.2 (2008): 335-396.

[3] X. Li, et al. "Development of ultra-short pulse reflectometry on the Experimental Advanced Superconducting Tokamak (EAST)." Review of Scientific Instruments 95, 073519 (2024).

[4] P. Sun, et al., "Millimeter-wave high-wavenumber scattering diagnostic developments on EAST and NSTX-U", Review of Scientific Instruments 95, 083553 (2024).

[5] Domier, C.W., Zhu, Y., Pereira, R.J., Steer-Furderer, J., Li, X., Chen, R., Xu, G., Sun, P. and Luhmann, N.C., 2022. Ultrashort Pulse Reflectometry (USPR) diagnostic for EAST. *Journal of Instrumentation*, *17*(02), p.C02010.

[6] G. Yu, et al., "Diagnosing the pedestal magnetic field and magnetohydrodynamics radial structure with pedestal–scrape of layer electron cyclotron emission radiation inversion in H-mode plasma." Review of Scientific Instruments 93.10 (2022).

[7] Y. Zhu, et al, 2020. W-band system-on-chip electron cyclotron emission imaging system on DIII-D. *Review of scientific instruments*, *91*(9).

[8] G. Yu, , et al. 2021. Noise suppression for MHD characterization with electron cyclotron emission imaging 1D technique. *Plasma Physics and Controlled Fusion*, *63*(5), p.055001.

[9] J. Chen, et al. "Faraday-effect polarimeter diagnostic for internal magnetic field fluctuation measurements in DIII-D." Review of Scientific Instruments 87.11 (2016).

[10] Y. Zhu, et al., "System-on-chip upgrade of millimeter-wave imaging diagnostics for fusion plasma." Review of Scientific Instruments 92.5 (2021).

[11] X. Li, et al. "System-on-chip Technology Application on Microwave Imaging Reflectometer." *Plasma Physics and Controlled Fusion* (2025).

[12] X. Li, et al., “GaN-based W-band receiver chip development for fusion plasma diagnostics.” Journal of Instrumentation 19 P06046 (2024).

[13] P. Sun, et al. "Development and Preliminary Results of 270 GHz Microwave Forward Scattering Diagnostic System on the Experimental Advanced Superconducting Tokamak (EAST)." Plasma Physics and Controlled Fusion (2025) 67 085029.

[14] X. Li, et al. "The Dual-Electron Cyclotron Emission Based Measurement of 3D Structures on DIII-D Tokamak." Plasma Physics and Controlled Fusion (2025).

[15] Y. Liu, et al. "Overview of the electron cyclotron emission measurements on EAST." Fusion Engineering and Design 136 (2018): 72-75.

[16] T. Rhodes, et al. "Design elements and first data from a new Doppler backscattering system on the MAST-U spherical tokamak." Review of Scientific Instruments 93.11 (2022).

[17] V. H. Hall-Chen, et al. "Beam model of Doppler backscattering." Plasma Physics and Controlled Fusion 64.9 (2022): 095002.

[18] Bruschi, A., et al. "Observation of short time-scale spectral emissions at millimeter wavelengths with the new CTS diagnostic on the FTU tokamak." Nuclear Fusion 57.7 (2017): 076004.

[19] M. Stejner, et al. "Measurements of plasma composition in the TEXTOR tokamak by collective Thomson scattering." Plasma Physics and Controlled Fusion 54.1 (2012): 015008.

[20] X. Li, et al , 2021. High level of integration of front-end imaging optics system for electron cyclotron emission imaging diagnostics on the DIII-D tokamak. Fusion Engineering and Design, 172, p.112915.

[21] G. Yu, et al. "Modeling the electron cyclotron emission radiation signature from suprathermal electrons in a tokamak." Review of Scientific Instruments 95, 073505 (2024).

[22] X. Yu et al, “Numerical simulation of ultrashort-pulse reflectometry (USPR) on EAST.” Plasma Sci. Technol., 25 (2023) 125601.

[23] G. Yu, et al. "Modelling of the electron cyclotron emission burst from a laboratory tokamak plasma with loss-cone maser instability." *Journal of Plasma Physics* 90.6 (2024): 985900601.

[24] W. Choi, et al, 2018. Feedforward and feedback control of locked mode phase and rotation in DIII-D with application to modulated ECCD experiments. *Nuclear Fusion*, *58*(3), p.036022.

[25] T. Ihli, et al. "Review of blanket designs for advanced fusion reactors." Fusion Engineering and Design 83.7-9 (2008): 912-919.

[26] Y. Zhu, et al. "Millimeter-wave imaging diagnostics systems on the EAST tokamak." Review of Scientific Instruments 87.11 (2016).

[27] C. Qu, et al. "The co-located arrangement of ECEI and MIR microwave imaging diagnostics on EAST tokamak." Journal of Instrumentation 18.10 (2023): C10018.

[28] Y. Liang, et al. "Overview of the recent experimental research on the J-TEXT tokamak." Nuclear Fusion 59.11 (2019): 112016.

[29] X. Ren, et al. "Microwave Reflectometer for Density Measurement on J-TEXT Tokamak." 2020 IEEE 1st China International Youth Conference on Electrical Engineering (CIYCEE). IEEE, 2020.

[30] Z. Yang, et al. "Electron cyclotron emission radiometer upgrade on the J-TEXT tokamak." Review of Scientific Instruments 87.11 (2016).

[31] S. Qiu, et al., " Design of a 140 GHz waveguide notch filter for millimeter-wave receiver module protection in fusion plasma diagnostics.", Rev. Sci. Instrum. 95, 023503 (2024).

[32] L. Himes, et al., “The 140 GHz notch filter development for millimeter-wave diagnostics protection on the stellarator Wendelstein 7-X.” Journal of Instrumentation 19 P10024 (2024).

[33] Y. Jiang, et al., W-band E-plane waveguide filter using compact beeline resonator, IEEE Microw. Wirel. Compon. Lett. 32 (4) (2022) 289–292.

[34] D. Pozar. Microwave engineering: theory and techniques. John wiley & sons, 2021.

[35] G. L. Matthaei, I. Young, and E. M. T. Jones, Microwave Filters, Impedance-Matching Networks, and Coupling Structures (Dedham, Mass. : Artech House, 1980).

[36] Dassault Systèmes, 2024. CST Studio Suite. Available at: <https://www.3ds.com/products-services/simulia/products/cst-studio-suite/>.